# Thermal performance and environmental impact of a wood frame wall enhanced with clay-based plaster: an experimental and simulation analysis

Z BOUTAYEB*, M LABAT, C OMS, S GINESTET

[1]LMDC, INSA/UPS Génie Civil, 135 Avenue de Rangueil, 31077 Toulouse cedex 04 France.

* Corresponding author boutayeb@insa-toulouse.fr

**Abstract.** This study evaluates the thermal performance and environmental impact of wooden frame walls with clay-based plaster by using the complementary properties of both materials. Clay's moderate density and high thermal capacity enhance the thermal behaviour of wood structures, making it a valuable addition to lightweight constructions. At the material scale, the thermal properties of clay-based plaster were experimentally characterized and integrated into thermal dynamic simulations to assess the influence of internal and external wall configurations on the indoor temperature in the framework of overheating in summer. A life cycle assessment was conducted to compare the environmental impact of three walls: one made of concrete, a wooden frame wall, and another using wood frame wall with clay-based plaster.

## 1. Introduction

The global urgency to address energy efficiency and environmental sustainability has placed the building sector under intense focus, prompting a re-evaluation of current practices and the exploration of new solutions. Traditional construction methods, often reliant on energy intensive materials like concrete, are increasingly being replaced by alternatives that prioritize natural, renewable, and recyclable resources [1]. Among these, wood and clay have gained significant attention due to their durability, low environmental impact, and favourable thermal properties. When combined, these materials offer a promising approach to meeting modern energy and environmental standards in construction.

Wooden structures, particularly in vertical extensions, provide a lightweight and sustainable alternative that meets structural and environmental requirements. However, the low thermal inertia of wood, resulting from its minimal mass, poses challenges in maintaining stable indoor temperatures. This limitation is especially pronounced in summer, when temperature fluctuations can affect indoor comfort. To reduce this effect, integrating materials with higher thermal capacity, such as clay-based plaster, presents a viable strategy for improving the thermal performance of wood-based buildings [2].

Clay, a material with a rich history in construction, has regained popularity in modern sustainable building practices. Its moderate density and relatively high thermal effusivity make it an effective complement to lightweight wooden structures, enhancing their thermal inertia and improving overall energy performance. Combining the benefits of wood and clay enables the development of construction systems that optimize thermal performance while reducing environmental impact.

This study explores the potential advantages of using modified wood-frame walls enhanced with clay-based plaster. Through experimental research and dynamic building simulations, it evaluates the thermal properties of these materials, the impact of internal and external wall configurations on indoor temperature stability, and the overall environmental performance of such systems. Additionally, a life cycle analysis (LCA) compares the environmental impact of wooden walls with clay-based plaster to that of conventional concrete walls. These findings provide valuable insights into how this approach advances sustainable building practices by reducing carbon emissions and improving energy efficiency.

## 2. Materials scale study: characterization of clay plaster

This section outlines the thermal properties of clay-based plaster at the material scale, focusing on parameters such as thermal conductivity, effusivity, density, and specific heat capacity. Various measurement techniques, including the graduated hot plate, hot wire, and differential scanning calorimetry (DSC), were employed, and the results were compared with existing literature.

*2.1. Components of material:*
Clay plasters are composed of a complex mixture of minerals [3]. The clay used in this study originates from the South-West of France, a region known for its rich kaolin deposits and high quartz content [4]. According to the manufacturer's specifications, the clay plaster is composed of 20% clay and 80% sand. X-ray diffraction (XRD) analysis confirmed the composition of the mixture, specifically identifying kaolinite and quartz, in agreement with the manufacturer's specifications.

*2.2. Thermal conductivity and Density:*
Thermal conductivity ($W \cdot m^{-1} K^{-1}$), which quantifies a material's ability to transfer heat under a temperature gradient, was measured in this study using the guarded hot plate and hot wire methods.

- **Guarded Hot Plate Method:** Samples (150 mm × 150 mm) were prepared according to the NF EN 12664 standard. Each sample was placed between two plates maintained at different temperatures ($\Delta T = 10$ K), and the heat flux was measured at average temperatures of 10°C and 23°C. Each measurement was repeated three times to ensure accuracy, and the average value was considered for analysis.
- **Hot Wire Method:** A wire was placed between two identical samples. By electrically heating the wire and monitoring the temperature rise over time, a rapid measurement of thermal conductivity was obtained, allowing direct comparison with the guarded hot plate results.

**Density Measurement:** The density of the clay plaster was determined using the hydrostatic weighing method with Garosolve to obtain the absolute density. The sample was ground into a fine powder (<80 µm) before measurement. The apparent density was measured using hydrostatic weighing in water. The average absolute density of the clay plaster was found to be 2.61 g.cm$^{-3}$, while the apparent density was 1.85 g.cm$^{-3}$.

*2.3. Effusivity and specific heat capacity:*
Thermal effusivity, which indicates a material's ability to exchange thermal energy with its surroundings, plays an important role in its thermal performance [2]. Additionally, [5] discusses the significant role of high thermal effusivity in earth bricks, which notably contributes to their thermal inertia. This property is essential for improving both thermal comfort and energy efficiency within buildings.

In this study, the hot plate technique [6] using the DesproTherm device was applied to estimate the thermal effusivity of the clay plaster. Results indicated an effusivity of approximately 1160 J.m$^{-2}$. K$^{-1}$. s$^{-1/2}$.

To determine the specific heat capacity ($J \cdot kg^{-1} \cdot K^{-1}$), differential scanning calorimetry (DSC) was performed in accordance with ASTM E1269 standard. Dried samples were heated at a rate of 10°C.min$^{-1}$ from 10°C to 60°C, and the specific heat capacity was derived from the resulting heat flow curve.

*2.4. Experimental results:*

The following figures illustrate the relationships between thermal conductivity, specific heat capacity, and density for the tested clay plaster, alongside reference values from the literature.

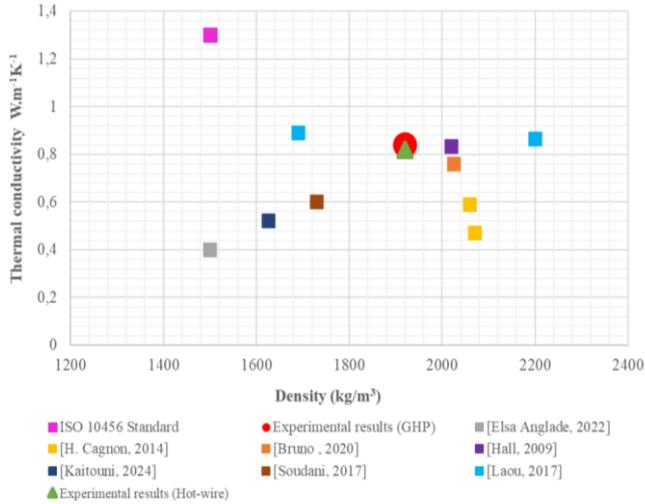

**Figure 1:** Thermal Conductivity vs. Density of Clay-Based Plaster **[5] [7] [8] [9] [10] [11] [12]**

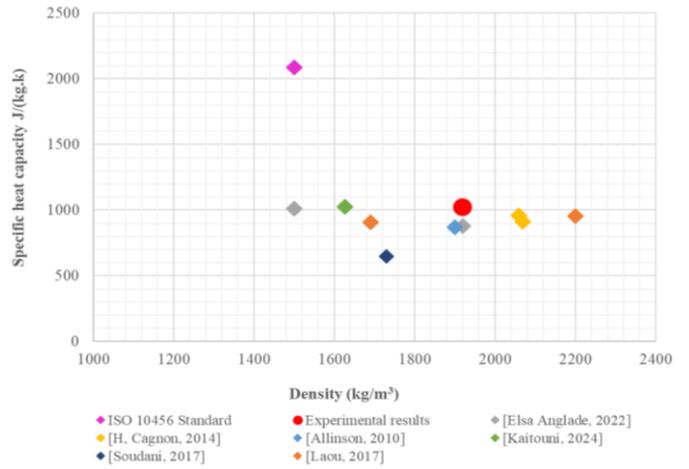

**Figure 2:** Specific Heat Capacity vs. Density of Clay-Based Plaster **[5] [10] [13] [12] [9] [8]**

The results confirm a clear relationship between density and thermal conductivity, in line with previous research. The clay plaster has a thermal conductivity of approximately 0.82 W.m$^{-1}$K$^{-1}$ and a specific heat capacity of around 1124 J. kg$^{-1}$.K$^{-1}$, which are consistent with existing studies. The slight deviations observed in certain data points could be due to compositional variations, or measurement technique differences. Despite these minor variations, the overall agreement with literature values validates the reliability of the experimental methods used in this study.

These experimentally derived properties are essential for thermal modelling. The following section presents numerical simulations that integrate these values to evaluate the thermal performance of wood-frame walls incorporating clay-based plasters under real building conditions.

**3. Dynamic building simulations:**
To assess the thermal performance of integrated wood and clay construction systems, dynamic simulations were conducted using Pleiades Comfie [7], a French software that complies with current national and European standards. This software enables dynamic modelling of building energy performance, allowing to determine the indoor temperature variation under realistic solicitations. The simulation, part of the CECOB project (Summer Comfort and Timber Construction), was based on a real residential building in Paris, consisting of five floors: a ground floor, two concrete floors, and two additional wooden heightening floors. In the baseline configuration, the internal wall surface consists of two layers of gypsum plasterboard (BA18), in line with conventional light wood frame construction. In the clay-based plaster configuration, the gypsum boards are replaced with a single layer of internal clay plaster, while the rest of the wall build-up remains unchanged.

To evaluate the impact of the wall's thermal inertia on the indoor conditions, three construction scenarios were analysed based on the effective heat capacity of vertical walls. The calculation method of the effective heat capacity used in this study is adapted from the approach described by [14], which is based on the specific heat capacity ($C_i$), density ($\rho_i$) of each layer, and thermal penetration depth ($\delta_i$), calculated for a 24 hour excitation period. Instead of considering the full material thickness, only the thermally active portion contributes to heat storage, ensuring a realistic representation of building thermal response. First, the effective heat capacity of each vertical wall was determined using:

$$C_{eff} = \sum(C_i \times \rho_i \times \delta_i) \qquad [\text{J} \cdot \text{m}^{-2} \cdot \text{K}^{-1}] \tag{1}$$

To obtain the room-level effective heat capacity for each apartment, the sum of all contributing walls is normalized by the total floor area:

$$C_{R,eff} = \frac{\sum(C_{eff,i} \times S_l)}{floor\ area} \qquad [\text{J} \cdot \text{m}^{-2} \cdot \text{K}^{-1}] \tag{2}$$

This normalization allows for a consistent comparison between apartments of different sizes and construction configurations. Figure 3 below presents the floor plan of the fourth floor, where the analysed apartments T1 and T4 are located. The table within the figure summarizes the effective heat capacity values for each construction scenario, allowing for a direct comparison between the two apartments. Each apartment (T1 and T4) was simulated as a single thermal zone, with the analysis focusing on the maximum recorded indoor temperature within the zone. Simulations were performed using weather data from Paris, focusing on the hottest week of the year to assess summer comfort and indoor temperature stability. All windows were assumed to be unshaded (no shutters), and apartments were considered occupied, with internal heat gains included.

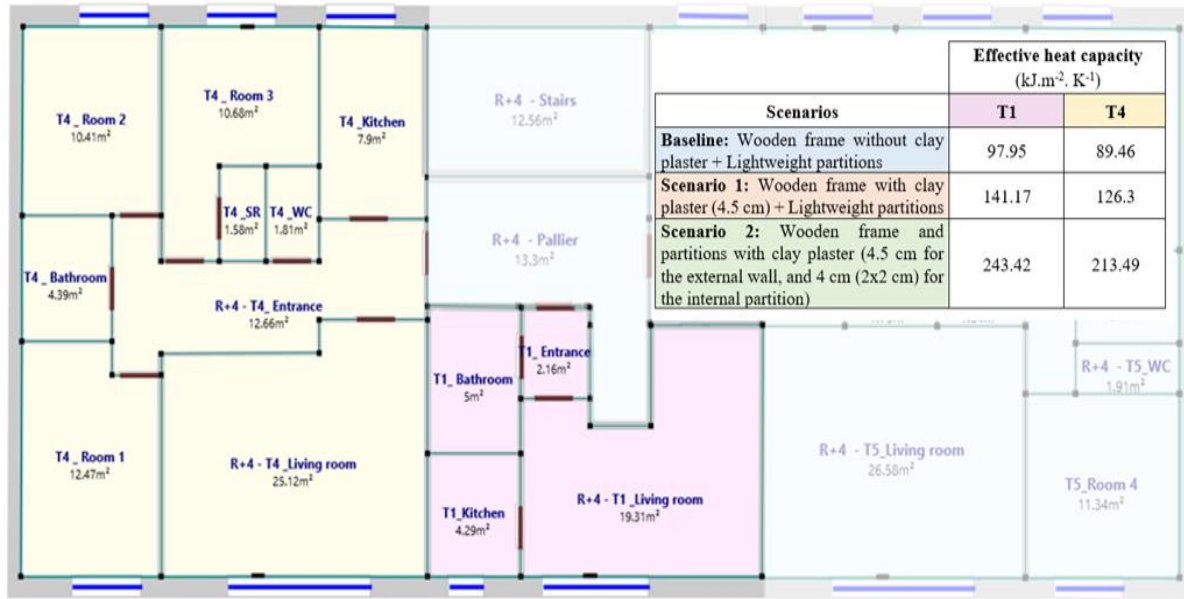

**Figure 3:** Fourth-floor plan and effective heat capacity of T1 and T4 apartments under different scenarios. The "Wooden frame" represents the main structural framework, while "Lightweight partitions" refer to internal walls separating rooms.

The results of the dynamic simulations are summarized in Figure 4, which illustrates the relationship between effective heat capacity and the maximum operative temperature for both apartments under different construction scenarios.

In the baseline configuration, the effective heat capacity of the T1 apartment is higher (97.95 kJ·m$^{-2}$·K$^{-1}$) than that of the larger T4 apartment (89.5 kJ·m$^{-2}$·K$^{-1}$). This is due to the interior partitions: both 100 mm and 160 mm thick partition walls were used, but the T1 apartment contains a greater proportion of 160 mm walls, which increases its total thermal mass. while T4 is mostly composed of 100 mm partitions.

However, the results shows that the maximum operative temperature in T4 is lower than in T1. This might be because the wall orientation is more varied for T4 than for T1. When clay plaster is added to the walls (scenario 1), the maximum temperature decreases but the difference exhibited between the two apartments is maintained: although the effective heat capacity in T1 remains higher than in T4, the temperature gain reduction is more pronounced in T4 (1.4 K) compared to T1 (1 K).

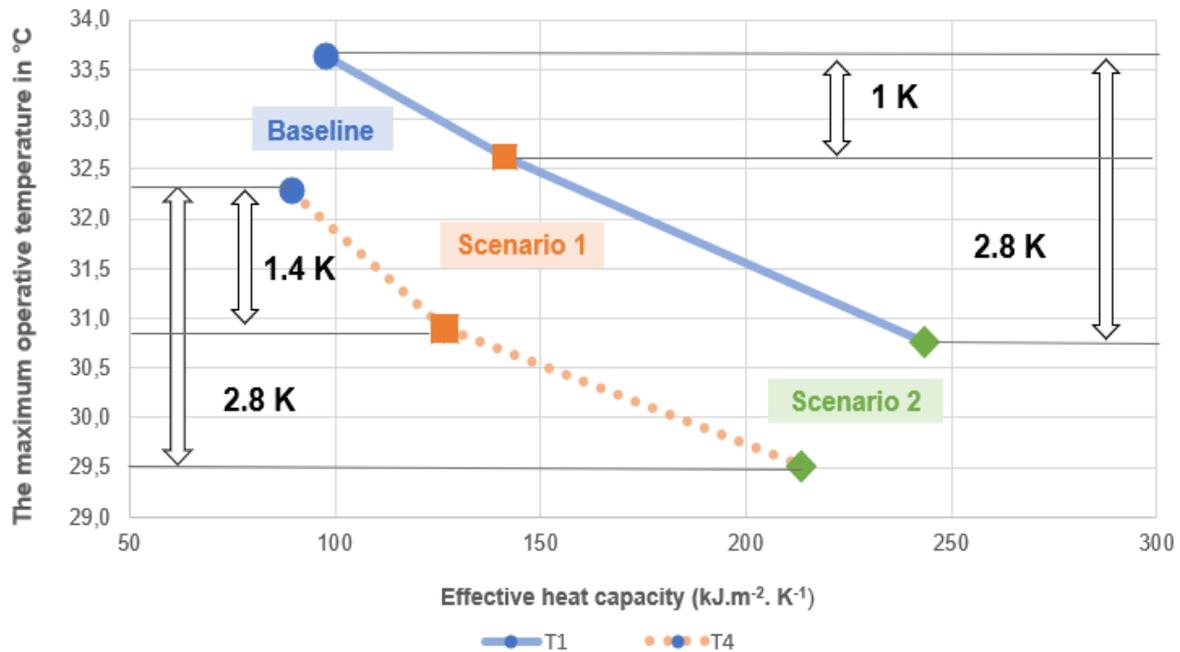

**Figure 4:** Impact of Effective Heat Capacity on Peak Operative Temperature for Two Apartment Types

When clay plaster is applied to all vertical surfaces (scenario 2), the reduction in peak temperature (ΔT = 2.8 K) is the same, yet effective heat capacity is higher in T1. This might suggest that T1 benefits from its higher thermal mass, while T4, with more varied wall orientations, may take advantage of a more distributed pattern of solar gains. The marked effectiveness of scenario 2 in limiting peak temperatures demonstrates clearly the benefit of integrating clay-based plaster on both façade walls and internal partitions, significantly reduces overheating risks.

To complement the thermal performance analysis, a Life Cycle Assessment (LCA) was conducted to evaluate and compare the environmental impacts of three wall configurations: a conventional concrete wall, a wood-frame wall with plasterboard, and a wood-frame wall with clay-based plaster.

**4. Life cycle assessment:**
Life cycle assessment is increasingly used to evaluate the environmental impact of construction materials and support sustainable decision-making. In this study, a comparative LCA was conducted to assess three wall configurations; concrete, wood frame with plasterboard, and wood frame with clay-based plaster, focusing on the production and end of life phases. The analysis was performed using SimaPro software[15] and the Ecoinvent database, and considered two key indicators: Global Warming Potential (GWP) and Cumulative Energy Demand (CED).

The results of the life cycle analysis (Figure 5) show significant differences between the three wall configurations in terms of cumulative energy demand (CED) and global warming potential (GWP). The concrete wall has the highest environmental impact, with non-renewable energy consumption reaching approximately 846 MJ.m$^{-2}$ and a GWP of 79.1 kg $CO_2$ eq.m$^{-2}$. This is mainly due to the energy-intensive production of concrete, particularly cement, and its low contribution from renewable sources.

In contrast, the two wood-frame wall configurations demonstrate considerably lower impacts in both energy and $CO_2$ emissions. The integration of clay-based plaster further reduces GWP from 32.5 to 29.6 kg $CO_2$ eq.m$^{-2}$ and increases the share of renewable energy use. This improvement is attributed to the low processing requirements and local availability of clay, which offsets the environmental impact of industrial materials. It is also notable that while the non-renewable energy demand of the two wood-based systems is relatively close (519 - 541 MJ.m$^{-2}$), the wood wall with clay contains a higher proportion of renewable raw materials and has a better carbon performance. These findings highlight the environmental benefits of using natural, bio-based, and low-impact materials like clay in combination with wood framing.

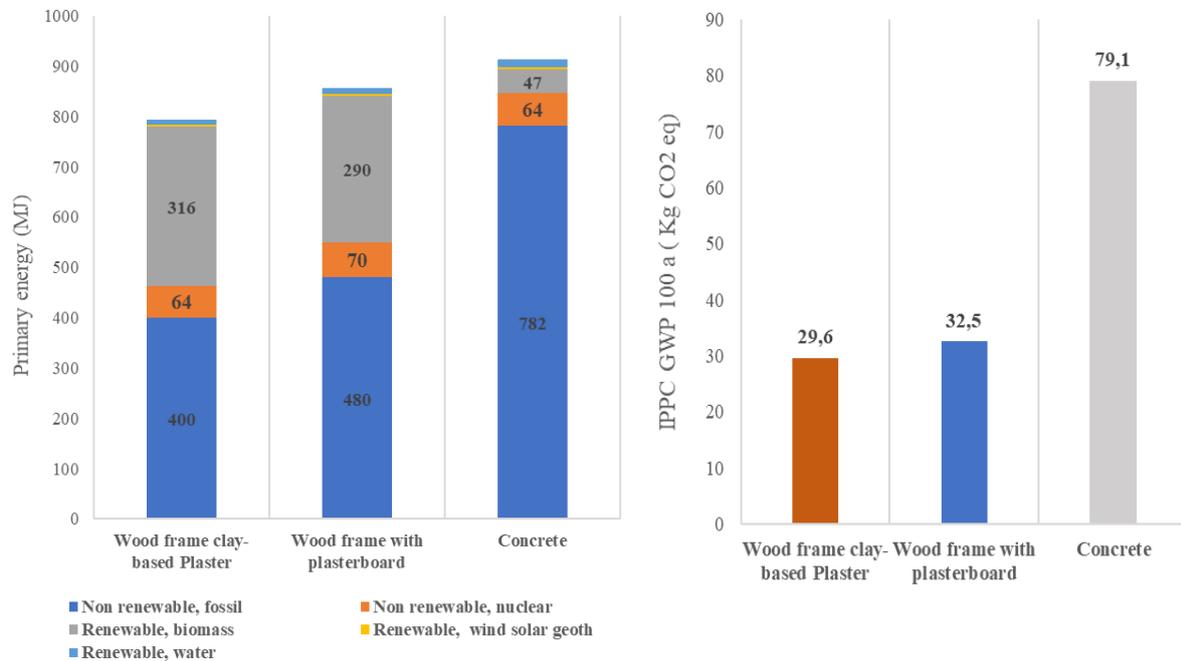

**Figure 5.** Environmental Performance of Wall Configurations: Cumulative Energy Demand and Global Warming Potential

Overall, the results emphasize the value of integrating clay-based materials into lightweight construction systems, not only for their thermal performance benefits but also for reducing life cycle energy use and greenhouse gas emissions.

## 5. Conclusions

This study highlights the dual benefit of integrating clay-based plaster into wood frame wall construction. On the thermal side, clay contributes to an increase in the effective heat capacity of wood-frame walls, which helps reduce indoor temperature peaks during summer conditions. In addition, the good performance of apartments with a transversal layout, like T4, suggests that the way spaces are arranged can be just as important for thermal comfort as the materials used.

From an environmental perspective, the life cycle assessment confirmed the advantages of clay-based plaster with wood frame walls over conventional concrete solutions, both in terms of reduced non-renewable energy demand and lower carbon emissions. Among the wood frame systems, the wall with clay-based plaster performed slightly better, particularly due to the low-impact, locally available nature of clay.

In summary, combining natural materials like clay and wood within well-designed layouts offers an effective strategy for reducing overheating risks and minimizing environmental impacts in lightweight construction systems.

To further refine these findings, ongoing full-scale wall tests are being conducted to assess the actual thermal performance of the wall systems and validate these preliminary results.

## 6. Acknowledgments:

This research was supported by the French Environment and Energy Management Agency (ADEME) as part of the doctoral funding program. The authors would also like to thank their institutional partners, including FCBA, ENSAP Bordeaux (GRECCAU), and CCM, for their valuable collaboration and technical support throughout the CECOB project.

## 7. References:

[1] B. P. Raj *et al.*, « A Review on Numerical Approach to Achieve Building Energy Efficiency for Energy, Economy and Environment (3E) Benefit », *Energies*, vol. 14, nº 15, p. 4487, juill. 2021, doi: 10.3390/en14154487.

[2] Luka Pajek *et al.*, « Improving thermal response of lightweight timber building envelopes during cooling season in three European locations », *Journal of Cleaner Production*, vol. 156, p. 939-952, juill. 2017, doi: 10.1016/j.jclepro.2017.04.098.

[3] J. Geng, « Effects of high temperature treatment on physical-thermal properties of clay », *Thermochimica Acta*, 2018.

[4] BRGM, « Kaolin et argiles kaoliniques », Service géologique national, févr. 2018.

[5] H. Cagnon, J. E. Aubert, M. Coutand, et C. Magniont, « Hygrothermal properties of earth bricks », *Energy and Buildings*, vol. 80, p. 208-217, sept. 2014, doi: 10.1016/j.enbuild.2014.05.024.

[6] N. Laaroussi, A. Cherki, M. Garoum, A. Khabbazi, et A. Feiz, « Thermal Properties of a Sample Prepared Using Mixtures of Clay Bricks », *Energy Procedia*, vol. 42, p. 337-346, 2013, doi: 10.1016/j.egypro.2013.11.034.

[7] A. W. Bruno, D. Gallipoli, C. Perlot, et H. Kallel, « Thermal performance of fired and unfired earth bricks walls », *Journal of Building Engineering*, vol. 28, p. 101017, mars 2020, doi: 10.1016/j.jobe.2019.101017.

[8] Elsa Anglade, « Évaluation des performances des matériaux en terre crue par homogénéisation poro-mécanique analytique non linéaire », 2022.

[9] S. Idrissi Kaitouni *et al.*, « Energy and hygrothermal performance investigation and enhancement of rammed earth buildings in hot climates: From material to field measurements », *Energy and Buildings*, vol. 315, p. 114325, juill. 2024, doi: 10.1016/j.enbuild.2024.114325.

[10] L. Soudani, M. Woloszyn, A. Fabbri, J.-C. Morel, et A.-C. Grillet, « Energy evaluation of rammed earth walls using long term in-situ measurements », *Solar Energy*, vol. 141, p. 70-80, janv. 2017, doi: 10.1016/j.solener.2016.11.002.

[11] M. Hall et D. Allinson, « Assessing the effects of soil grading on the moisture content-dependent thermal conductivity of stabilised rammed earth materials », *Applied Thermal Engineering*, vol. 29, nº 4, p. 740-747, mars 2009, doi: 10.1016/j.applthermaleng.2008.03.051.

[12] Lamyaa Laou, « Evaluation du comportement mécanique sous sollicitations thermo-hydriques d'un mur multimatériaux (bois, terre crue, liants minéraux) lors de sa construction et de son utilisation. », 2017.

[13] D. Allinson et M. Hall, « Hygrothermal analysis of a stabilised rammed earth test building in the UK », *Energy and Buildings*, vol. 42, nº 6, p. 845-852, juin 2010, doi: 10.1016/j.enbuild.2009.12.005.

[14] M. Němeček et M. Kalousek, « Influence of thermal storage mass on summer thermal stability in a passive wooden house in the Czech Republic », *Energy and Buildings*, vol. 107, p. 68-75, nov. 2015, doi: 10.1016/j.enbuild.2015.07.068.

[15] « SimaPro - Sustainability insights for informed changemakers », SimaPro. Consulté le: 31 mars 2025. [En ligne]. Disponible sur: https://simapro.com/